# Holographic imaging of unlabelled sperm cells for semen analysis: a review

*Giuseppe Di Caprio[1,2,*], Maria Antonietta Ferrara[1], Lisa Miccio[3], Francesco Merola[3], Pasquale Memmolo[3], Pietro Ferraro[3], and Giuseppe Coppola[1]*

[1] Institute for Microelectronics and Microsystems, Unit of Naples - National Research Council, Naples 80121, Italy
[2] Currently at the Rowland Institute at Harvard, Harvard University, Cambridge, MA, 02142, USA
[3] Institute "E. Caianiello" - National Research Council, Pozzuoli 80078, Italy

Male reproductive health in both humans and animals is an important research field in biological study. In order to characterize the morphology, the motility and the concentration of the sperm cells, which are the most important parameters to feature them, digital holography demonstrated to be an attractive technique. Indeed, it is a label-free, non-invasive and high-resolution method that enables the characterization of live specimen. The review is intended both for summarizing the state-of-art on the semen analysis and recent achievement obtained by means of digital holography and for exploring new possible applications of digital holography in this field.

## 1. Introduction

Semen analysis is widely used as diagnostic tool for assessing male reproductive health in both humans and animals. In particular, semen analysis in humans is mainly used for couple's infertility exploration or to confirm successful of male sterilization procedures. Moreover, infertile men also have a greatly increased risk of developing testicular cancer. So, whether or not a man intends having children, his fertility is a good indicator of his general health. On the other hand, in the zoo-technic field, animal semen analysis is commonly used in stud farming and farm animal breeding. The most important parameters measured in a semen analysis are the morphology, motility and the concentration of the sperm [1-10]. Thus, abnormal sperm features are the most common important indicators for male infertility. For this reason, there is growing interest in understanding both the spermatozoa morphological alterations and the kinematics/dynamics of the swimming spermatozoa. The main requirements for the new techniques that have been used to obtain such information are: the non-destructive method avoiding any alteration of the vitality of the analysed sperm, a label-free approach to reduce costs and exclude all adverse effects that may be introduced by labels and the independence on the experience of the technician and environmental conditions (such as, temperature, pH level, and duration). While spermatozoa are essentially transparent, and almost invisible when observed in optical bright-field microscopy, they have a different refractive index than the surrounding medium: the phase of the light transmitted by the sample registers this modulation in refractive index. A qualitative visualization of this phase contrast may be obtained by contrast interference microscopy (phase contrast or Nomarski/Zernicke interferential contrast microscopy). However, it is difficult and time-consuming to obtain a quantitative morphological imaging. In fact, a fine z-movement of the biological sample is required in order to acquire a collection of different planes in focus. This collection of acquired images is used in post-elaboration to produce a 3-D image of the object under investigation [11]. The same approach has been used to obtain information about sperm motility. Nevertheless, this two-dimensional intrinsic analysis implies a partial in-plane representation of the motility features due to difficulty to track the 3D spatial motion of spermatozoa that quickly move out of focus. Commercial computer-assisted semi-automated system (CASA) [12] provides in-plane information such as: straight-line velocity, curvilinear velocity and linearity. CASA is based on a combination of light microscopy and sophisticated computer software, often using negative phase-contrast. In order to overcome these intrinsic limitations several approaches have been recently developed. In this contest, the optical approaches are deeply investigated. This review tries to summarize the state-of-art on the semen analysis and recent achievement obtained by a Digital Holographic (DH) approach [13-16]. DH is a label-free, noncontact, non-invasive and high-resolution method that allows the recording and the numerical reconstruction of the phase and amplitude of the specimen's optical wavefront. Thus, 3-D quantitative sample imaging can be automatically produced by numerical refocusing of a 2-D image at different object planes without mechanical realigning the optical imaging system [17]. Consequently a volumetric field can be reconstructed by means of a single image (the hologram). This approach enables the characterization of live specimen, and on the other hand, allows both reducing the size of the mass storage devices required for image saving and achieving a fast image transfer. DH has been successfully applied for real-time 3-D metrology for studying microelectromechanical systems (MEMS) [18], vibration analysis [19], particles [20], recognition of bioorganisms [21-25], and nano sized particle detection [26]. Furthermore, DH may allow quantitatively retrieving, in far field region, the amplitude and phase of the wavefront interacting with the structures themselves [27-30]. In the following sections, we will show that the unique potentialities of the holographic imaging have been used to provide structural information on both the morphology and motility of sperm cells. The possibility to add the third dimension in the sperm analysis can provide a better understanding of the sperm behaviour and its relation with male infertility [31]. In Section 2, principles of operation of digital holography will be described. A review of morphological images obtained by the holographic approach is reported in Section 3, whereas in Section 4, the possibility to use the holographic imaging of spermatozoa inside microfluidic channel is described. Then in Section V, the holographic approach used to track the 3D spatial motion of spermatozoa is reported. Finally, future trends and conclusions are presented in last section.

## 2. Methods and Materials

### 2.a Principle of operation of digital holography

Biomedical holographic imaging of living cells is a fast on-going research field and several parts of the technology and applications have been recently reviewed in various articles and books [15, 16]. So, in this section we briefly describe the basic principles of the technique. An optical field consists of amplitude and phase distributions but all detectors (or recording materials) register intensity only. If two waves of the same frequency interfere, the resulting intensity distribution is temporally stable and depends on the phase difference $\Delta\varphi$. This phase variation incorporates information about the morphology of the object under investigation. Thus, the holographic approach employs the interference to codify the phase information (i.e. the 3D information) into a recordable intensity distribution [32]. In particular, a full or partial coherent laser source is split in two beams: a *reference beam* $(R(x,y) = |R(x,y)|e^{i\varphi_R(x,y)})$ and a beam that illuminates the biological object. This object scatters the incoming beam forming the *object beam:*

$$O(x,y) = |O(x,y)|e^{i\varphi_O(x,y)} \qquad (1)$$

where the phase $\varphi_O(x, y)$ depends on the refractive index and thickness of both the biological sample and the material containing the object itself [17]. The *hologram,* that is proportional to the intensity of the interference between the reference and object waves, is acquired by an image sensor (CCD or CMOS). A sketch of the afore-described configuration is shown in Fig. 1. According the angle ☐ between the reference and object beams either on-axis (☐☐☐) or off-axis configurations can be adopted.

The image reconstruction procedure allows retrieving a discrete version of the complex optical wavefront present on the surface of the specimen under test. This optical wavefront is obtained by a numerical back-propagation of the product between the recorded hologram and a replica of the reference beam. Actually, this product generates three diffraction terms: zero-th order, virtual image, and real image. In an on-axis configuration, the real and conjugate images are superimposed. Thus, their separation requires either a spatial or temporal phase-shifting methods, which adds to the complexity and capture time [13] unless a partially coherent light is adopted, that allows reducing the speckle and multi-reflection interference noise [33].

By introducing a small angular separation between the two interfering beams in off-axis configuration, a spatial separation between the real and conjugate images is obtained at the expense of suboptimal use of sensor space-bandwidth product. This separation allows selecting and retrieving the real image, that is an exact replica of the object wavefront.

Obviously, according the final experimental goal, each adopted set-up is designed with the right requirements: magnification and numerical apertures of the microscope objectives, size of the image sensor, the wavelength of the laser source, and so on. The object beam intensity is always set well below the level for causing any damage to the spermatozoa structure and functionality. Some of the results presented in the next paragraphs have been acquired using a variant of this method, called in-line holography [33], where the interference of two components of the same light beam is used to generate the hologram.

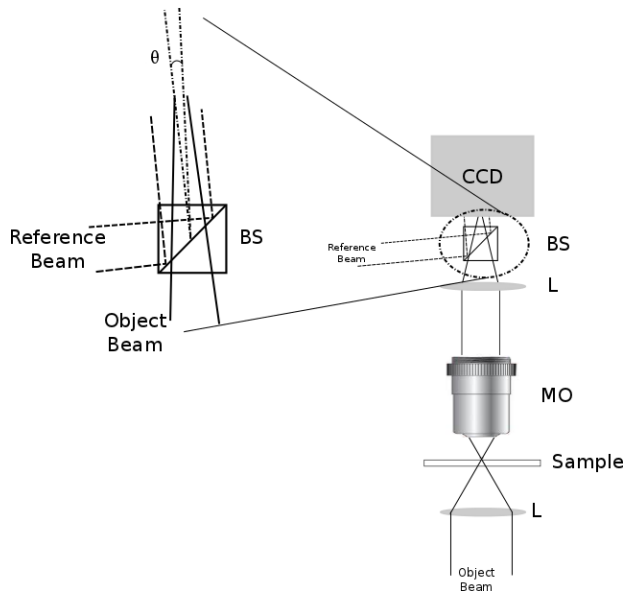

**Figure 1.** Sketch of the principle of hologram formation (L:lens, MO: Microscope Objective, BS: Beam Splitter).

Finally, the possibility offered by DH to manage the phase of the reconstructed image allows removing and/or compensating the unwanted wavefront variation (such as, optical aberrations, slide deformations) [34-36].

## 2.a Structure of a spermatozoon

A spermatozoon is a polarized motile cell, which delivers the haploid male genome to the oocyte, introduces the centrosome and triggers the oocyte egg into activity. In Figure 2, DIC image of a sperm is reported, where both the tail and the head are visible. The tail is composed of the neck, middle, principal and end pieces and is responsible of sperm motility, necessary to the spermatozoon to reach the oocyte [37]. The head contains three functional parts: 1) the nucleus with a haploid set of chromosomes, in which deoxyribonucleic acid (DNA) is packaged into a volume that is typically less than 10% of the volume of a somatic cell nucleus; 2) the acrosome, a large Golgi-derived secretory vesicle on the proximal hemisphere of the head containing an array of hydrolytic enzymes necessary for digesting the zona pellucida during penetration of the oocyte; 3) the perinuclear theca, a rigid capsule composed of disulfide bond-stabilized structural proteins amalgamated with various other protein molecules. Shape and size of spermatozoa vary by species and a number of studies indicate that sperm morphology best predicts of outcome for natural fertilization [4, 38], intra-uterine insemination conventional IVF and ICSI [39].

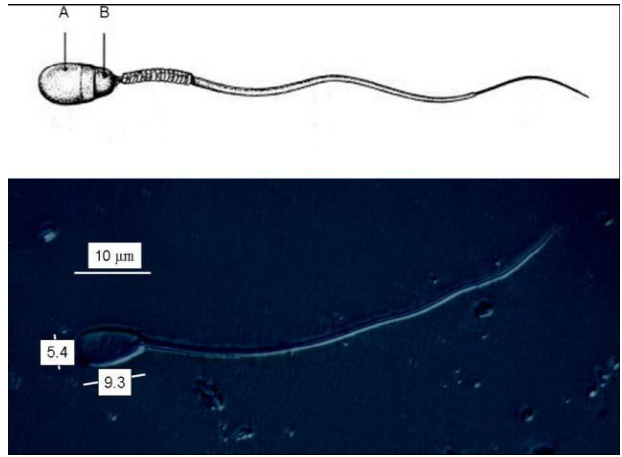

**Figure 2** (Upper side) Structure of the spermatozoon composed of head, neck, middle piece, tail and end piece. The male gamete's head is partially covered by acrosome (A) and the remaining part is named as postacrosomal region (B). (Lower side) Image of a bovine spermatozoon obtained by an optical-contrast interference microscope [46].

## 3. Holographic imaging of sperm cells

The first holographic image of a spermatozoon was provided by Mico *et al.* [40] in 2008. However, authors used the biological object just to demonstrate the effectiveness of their super-resolution imaging method. Actually, due both the appealing and application potentialities of sperm characterization, several research groups have used spermatozoa specimens as sample to show the effectiveness of their proposed method [41-45]. However, neither quantitative nor behaviour analysis were performed on the morphology and motility of sperm.

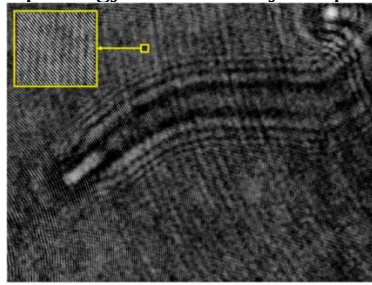

**Figure 3.** Acquired hologram, a region is enhanced in order to show the interference pattern (inset) [46].

The possibility to use a quantitative morphological analysis in the sperm analysis to provide a better understanding of the sperm behaviour was reported by Coppola's group [46]. Bovine sperm cells were considered in their experiments. In Figure 3, the image of an acquired hologram is reported, whereas its inset shows the intensity of the fringe pattern due to the superimposition of the object and reference beam.

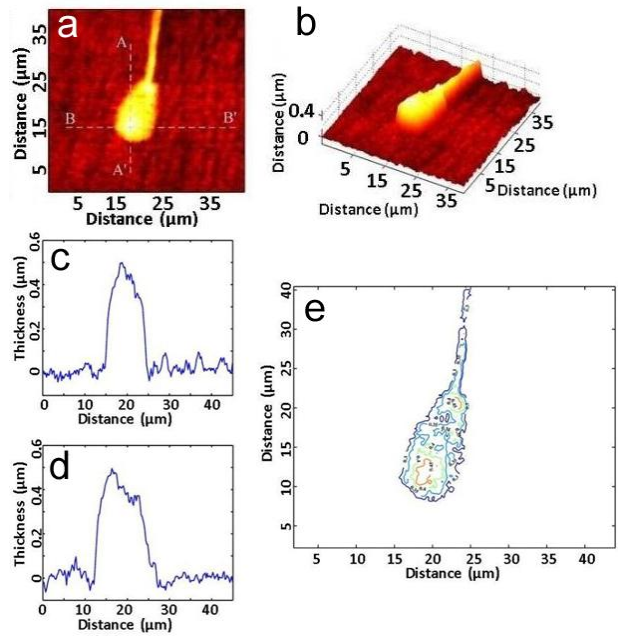

**Figure 4** (a) Pseudocolor plot of a phase-contrast map for a bovine spermatozoon. (b) Pseudo 3-D representation of the thickness of a bovine spermatozoon. (c) and (d) profiles plot along the lines AA' and BB'. (e) Isolines relative to five different thicknesses of the reconstructed image [46].

In Figure 4(a), a pseudocolor plot of the unwrapped phase-contrast map reconstruction of a bovine spermatozoon is reported. For the reported analysis, the spermatozoa were fixed and without surrounding liquid. Figure 4(b) illustrates the quantitative reconstructed morphology. Therefore, a refractive index for the sperm cell of 1.35 is estimated [46].
The great advantage to manage quantitative information allows carrying out different numerical analysis; such as: estimation area, profiles along particular directions or selection of different zone. In particular, in Figure 4(c) and (d) are reported the quantitative profiles of the spermatozoon morphology along the lines AA' and BB' illustrated in Figure 4(a), respectively. Moreover, in Figure 4(e), as example, isolines relative to five different thicknesses of the sample (i.e. [0.30, 0.35, 0.40, 0.45, 0.50] µm) are visualized. Then, for each region defined by the isolines, the occupied area and the relative volume have been numerically estimated and the carried out values are reported in Tab. I.

**Table I.** Numerical estimation of the area and volume of the regions defined by the isolines displayed in Fig.4 [46].

| Reference thickness of isoline [µm] | Calculated area [µm$^2$] | Calculated volume [µm$^3$] |
|---|---|---|
| 0.30 | 73.7 | 27.9 |
| 0.35 | 51.8 | 20.7 |
| 0.40 | 24.4 | 10.5 |
| 0.45 | 3.5 | 1.62 |

The holographic imaging has been used to visualize the morphology of abnormal sperm, too. In particular, in Figure 5(a) the reconstructed image of a spermatozoon with a cytoplasmatic droplet along the tail is reported. Cytoplasm surrounding the sperm cell is accumulated during maturation and in the last phases of the process is extruded from the cell. However, cytoplasmatic residues may persist in the cell as a droplet and are retained in the tail. When droplet is located in the neck region this defect is defined as "proximal droplet", while when the defect involves the middle piece is known as "distal droplet" [37]. Thus, the presence of drops along the tail is connected to the degree of cell maturation and may indicate an excessive utilization of a donor. In Figure 5(b) a reconstructed image with a bent tail sperm cell is displayed. This defect can be associated with low sperm motility. This defect, if it is present in the semen either before or after the freezing process, underlines a reproductive problem of the donor. When this anomaly appears with high frequency only in frozen semen, it can indicate that the spermatozoa have been subjected to hypo-osmotic stress possibly due to an improper use of freezing extender and to an extremely low concentration of solutes [37]. In Figure 5(c) a sperm with broken acrosome is displayed. This defect is not common in fresh semen but it can be present with high percentage in frozen semen samples. The loss of acrosomal substances indicates premature acrosome activation far from the site of fertilization. This defect is usually due to incorrect sperm handling during the freezing process.

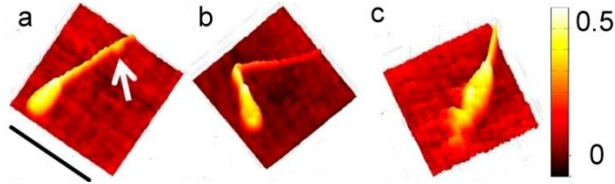

**Figure 5.** Pseudo 3D representation of the thickness of a spermatozoon with: (a) a cytoplasmatic droplet along the tail; (b) a bent tail; (c) an acrosome broken. Scalebar is 40 ?m, Colorbar is in ?m [46].

The possibility to quantitatively evaluate these kinds of measurements can add supplemental information to the bi-dimensional data obtained by the traditional optical microscopy in order to improve the understanding of the relationship between the abnormal morphology and the male infertility. In pursuit of this aim, the holographic approach has been also used to investigate the human sperm characteristics [47, 48]. Crha *et al.* [47] compared quantitative phase contrast of sperm heads in normo-zoospermia (NZ) and oligoasthenoterato-zoospermia (OAT). To individuate phase shift of sperms of about 3000 tested sperm cells, a detailed statistical analysis based on mean, median, standard deviation (SD) and confidence intervals (CI) were used. Table II summarizes the descriptive statistics of the phase shift according to NZ/OAT groups, whereas an example of sperms different phase shift is illustrated in Fig. 6.

**Table II.** Descriptive statistics for statistically significant ($\rho < 0.001$) phase shifts according to the NZ/OAT group [47].

| Sperm Group | Median | Mean | SD | CI |
| --- | --- | --- | --- | --- |
| NZ | 2.90 | 2.91 | 0.61 | 2.94 |
| OAT | 2.00 | 2.10 | 0.38 | 2.13 |

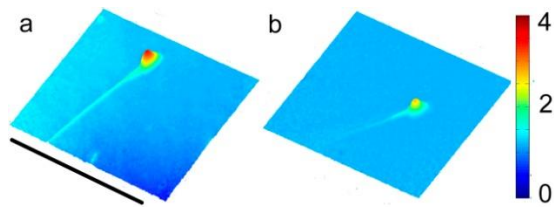

**Figure 6.** Three Dimensional Holographic images of the human sperm with different phase shifts: maximal phase shift 3.5 rad (a) and 2.45 rad (b). Scalebar is 60 ?m, colorbar is in rad [47].

According authors, the origin of the estimated difference of maximum phase shift of spermatozoa heads in NZ/OAT groups could be due to vary characteristics, such as: structural organization of the sperm DNA, condensation of the chromatin, alteration of protamines, histones or other proteins [49-51]. Nevertheless, the inhomogeneity of the background, i.e. visible in Fig 6a, might affect the phase shift evaluation.

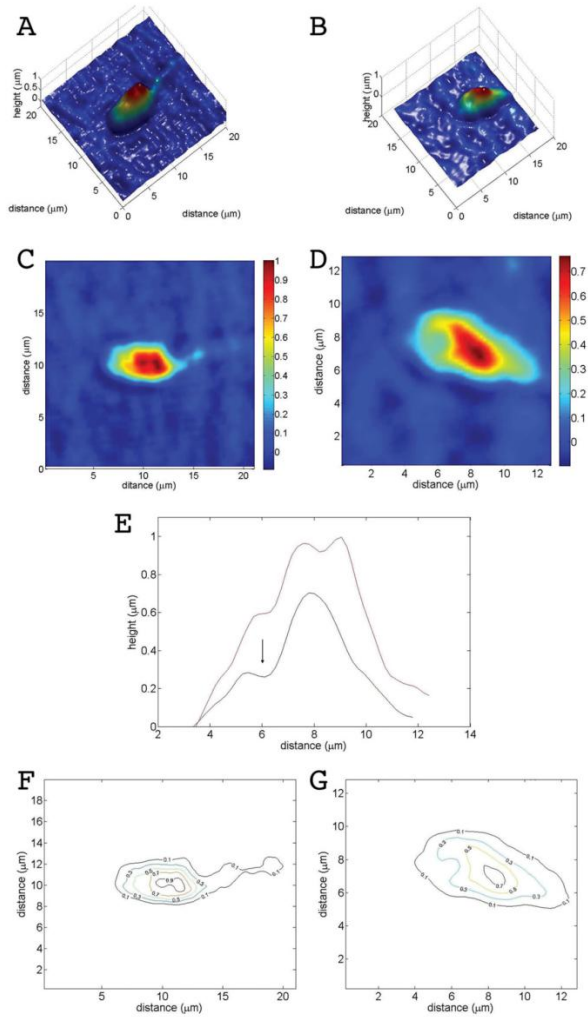

**Figure 7.** Comparison between a defect-free spermatozoon (left column) and a spermatozoon with vacuoles (right column). Results for a defect-free spermatozoon are reported as a quantitative three-dimensional representation (A), phase-contrast map (C), or as an isolines plot (F). For a spermatozoon with vacuoles, defects are reported as a quantitative three-dimensional representation (B), phase-contrast map (D), or as an isolines plot (G). (E) Comparison between the profile of the spermatozoon with vacuoles (curve indicated by arrow) and one without defects. [48]

The Dale's group [48] performed a comparison of 200 spermatozoa characterized both by a semi-automated digitally enhanced Nomarski microscopy (DESA) and by the Holographic Imaging. In particular, morphometrical, morphological and volumetric measurements have been evaluated on normal and vacuolated human spermatozoa. The main motivation for this kind of analysis was the great influence of nuclear vacuoles in the sperm head on the fertilization capacitance of sperm cell [8]. Generally, these anomalies has been analysed only from the two-dimensional point-of-view. The predominant locations of the vacuoles are the apical region and the acrosome-postacrosomal sheath junction, but they have also been found throughout the sperm head [52]. The authors aimed to add three-dimensional information, too. In particular, they carried out on the same spermatozoa both the DESA and Holographic analysis. In order to be sure to analyse the same spermatozoa with both approaches, a grid of 20x20 circles (with a radius of 100 µm) was placed over the microscope slide. Only the spermatozoa pinpointed inside the circles were characterized. In Table III mean morphometric values of normal sperm heads obtained by DESA and Holographic techniques are summarized.

**Table III.** Mean morphometric values of normal human sperm heads obtained by DESA and Holographic techniques [48].

| Imaging | Length [μm] | Width [μm] | Perimeter [μm] | Area [μm²] |
|---|---|---|---|---|
| DESA | 5.1±0.6 | 3.5±0.4 | 13.8±1.4 | 14.1±2.0 |
| Holography | 5.6±0.3 | 2.9±0.5 | 14.3±1.2 | 13.0±1.2 |

The morphometric values of the analysed sperm cells are consistent within the uncertainty. However, only the holographic approach can provide volume estimation. In Figure 7 a quantitative comparison carried out by the authors between a "normal" human spermatozoon and a spermatozoon with vacuoles is shown.
Table IV shows three distinct groups of spermatozoa defined using two morphometric variables: head length and head width. Mean values of the total volume of the spermatozoa minus the vacuoles volume are also reported.

**Table IV.** Mean volumetric values of vacuolated sperm clustered in three different subpopulations [48].

| | Volume [μm³] | |
|---|---|---|
| **Sperm dimensions** | Total | Total-Vacuoles |
| length<2.9μm, width<4.2μm | 5.8±0.7 | 4.0±0.8 |
| 2.9 < length < 3.7μm; 4.2 < width < 5.3 μm | 8.2±0.8 | 6.4±0.8 |
| length > 3.7 μm, width > 5.3μm | 10.1±0.8 | 8.4±0.8 |

In literature, nuclear vacuoles have been described either as a crater defect in the spermatozoa of stallion [53] or as a pouch formation [54]. Results provided by the holographic approach show that spermatozoa with vacuoles had a reduced volume, and this reduction could be probably due to variation of the inner structure of the sperm head with loss of material.

## 4. Sperm cells flowing in a microchannel

The analyses carried out in the previous section are relative to spermatozoa fixed on glass slide for optical microscope. However, DH provides the advantage of directly observing spermatozoa in their native environment. For this aim, microfluidic systems have been developed [55]. In fact, complex and precise operation of sperm cells and minim liquid can be achieved by justified design of microfluidic structures [56]. The first holographic image of a spermatozoon inside a Polydimethylsiloxane-based microchannel is reported in [46]. The similar value of the refractive index of the sperm cell and the surrounding medium has not allowed a clear reconstruction of the analysed object (see Fig. 8).

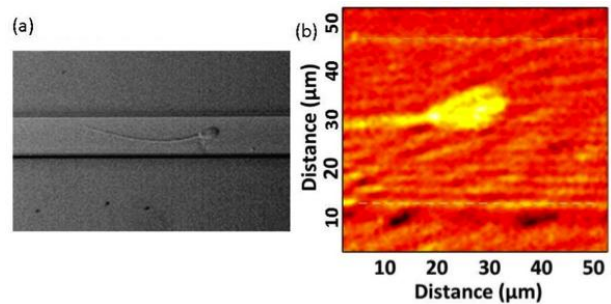

**Figure 8.** (a) DIC image of a spermatozoon into a microchannel. (b) Reconstruction image of a spermatozoon [46].

In order to obtain a more accurate 3D reconstruction of sperm cells into microchannel, Ferraro's group combined optical trapping capabilities with DH [57]. In particular, the authors used a laser to trap and induce rotation in a spermatozoon cell, meanwhile recording digital holograms of the specimen at different angles. 2D quantitative information from all the recorded holograms was then converted in a 3D rendering of the investigated cells. The contact-less and all optical proposed approach allows avoiding any mechanical sample and/or camera rotations typical of the tomographic microscopes [58, 59], furnishing a complete 3D visualization of the sample. In Fig. 9(a) a sketch of the interaction between the sperm cell and the trapping laser is reported. This interaction allows putting in rotation the cell and holographic images of different parts of the cell can be recorded without mechanical control. The rotating angular speed of the spermatozoa can be regulated as it depends on the power of the trapping beam. The bigger is the trap intensity, the faster is the rotation speed. The holographic images are combined to obtain a highly detailed 3D image (Fig. 9(b)). In ref. [57] the authors used the afore-described approach to evaluate the biovolume of spermatozoa (about 55 $\mu m^3$) providing a level of statistical significance greater than a standard 2D-based method. Through further implementation, the set-up could give the possibility to perform this "trapping and analysis" on multiple motile cells and almost in real-time.

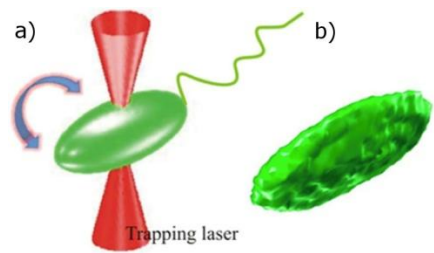

**Figure 9.** (a) Schematic interaction between the sperm cell and the trapping laser. (b) 3D reconstructions of spermatozoon head taken from different points of view [57].

## 5. Holographic tracking of sperm cells

One of the main advantages of the holographic method is the possibility to retrieve a 3D quantitative sample imaging produced by numerical refocusing of a 2-D image at different object planes without realigning of the optical imaging system with mechanical translation. This aspect enables the characterization of live specimen [60], and in particular, to track the 3D spatial motion of spermatozoa that quickly move out of focus. The tracking approach allows retrieving many quantitative parameters useful for a semen analysis. In particular, to provide quantitative motion values, several parameters have been generally estimated [12], such as: curvilinear velocity (VCL), straight-line velocity (VSL), and average path velocity (VAP), linearity, and so on. The VCL refers to the total distance that the sperm head covers in the observation period and it is always the highest of the 3 velocity values. The VSL is determined from the straight-line distance between the first and last points of the trajectory and gives the net space gain in the observation period. This is always the lowest of the 3 velocity values for any spermatozoon. The VAP is the distance the spermatozoon has travelled in the average direction of movement in the observation period. Further, to describe the trajectory, velocity ratio values are often evaluated [12]. These are linearity (LIN), a comparison of the straight-line and curvilinear paths, wobble (WOB), a comparison of the average and curvilinear paths and/or Amplitude of lateral displacement (ALH). However, this information is provided as in-plane parameters. The unique potentialities of the holographic imaging allow adding 3-dimensional information about the trajectory followed by the sperm cells in a volume. This possibility can provide a better understanding of the sperm behaviour and its relation with male infertility [31].

Ozcan's group used a lensfree holographic imaging approach to dynamically track the 3D trajectories of human sperms across a large volume with submicron positioning accuracy [61]. This imaging setup is based on a method called in-line holography [33], where the interference of two components of the same light beam is used to generate the hologram. Authors used two partially-coherent LEDs (light-emitting-diodes) at two different wavelengths that simultaneously illuminate the sperms at two different angles (red at 0° and blue at 45°). In Fig. 10 the set-up of the developed system is sketched.

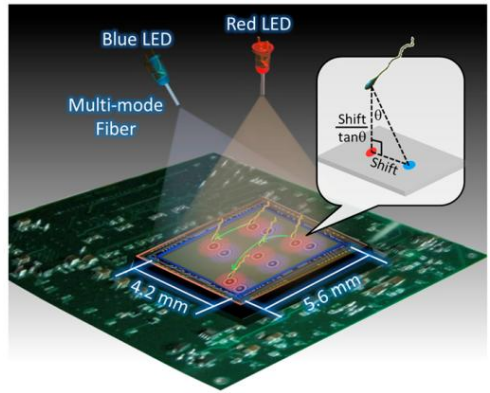

**Figure 10.** A schematic view of the imaging system [61].

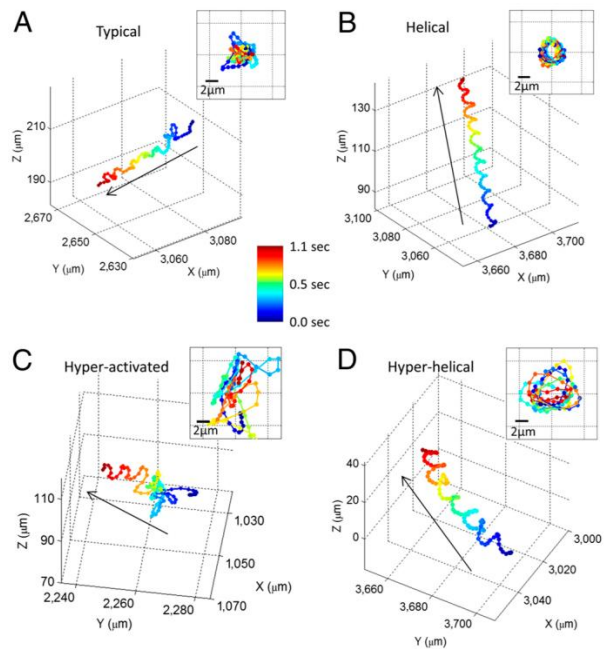

**Figure 11.** Swimming patterns of human sperms observed through the on-chip lensless holographic microscope of Ozcan's group. (A) The typical pattern. (B) The helical pattern. (C) The hyperactivated pattern. (D) The hyperhelical pattern. The inset in each panel represents the front view of the straightened trajectory of the sperm [61].

The 3D location of each sperm is determined by the combination of the images reconstructed in the vertical (red) and oblique (blue) channels. By means of this approach authors successfully tracked the 3D dynamic swimming patterns of human sperm across a field-of-view of >17 mm² and depth-of-field of about 0.5–1 mm. In particular, the developed method allowed observing four major categories for the swimming patterns. The most part of sperm (90%) moves forward swiftly along a slightly curved axis. Then, approximately 4-5% of motile human sperms move with a helical trajectory with a noticeable movement along the z-axis. Less than 3% of sperms exhibit a hyperactivated 3D swimming with large lateral movements. Finally, about the 0.5% of motile human sperms were characterized by a hyperhelical pattern. The evaluated patterns are displayed in Fig. 11.

By means of the observation of these trajectories on a large number of sperm cells (> 1,500) a statistic analysis on various parameters has been estimated; in particular, in Table V some of these parameters are summarized.

**Table V.** Mean values of some parameters related to the motility of human sperm: Curvilinean velocity (VCL), Straight-line velocity (VSL), Linearity, Amplitude of lateral head displacement (ALH) [61].

|  | VCL [μm /sec] | VSL [μm /sec] | Linearity [μm/μm] | ALH [μm] |
|---|---|---|---|---|
| Mean value | 88.0±28.7 | 55.7±24.9 | 0.61±0.21 | 5.4±2.9 |

Furthermore, thanks to the high accuracy of the technique, the authors observed that among the helical human sperms, a significant majority (approximately 90%) preferred right-handed helices over left-handed ones, with a helix radius of approximately 0.5–3 μm. Nevertheless this procedure, because of the low spatial resolution, is incapable of imaging single cell features.

Di Caprio *et al.* [62] have proposed a different approach that allows both the three-dimensional tracking and an imaging of single human sperm cell. In particular, the authors used an off-axis set-up and the capabilities of holographic technique to resolve in-focus amplitude and phase maps of the sperm cells, independently of focal plane of the sample image. In particular, in order to estimate the sperm motility, a set of holograms at a constant sample – microscope objective distance were acquired. Each retrieved phase map is used to evaluate the X and Y coordinates of sperm cells by means of a shape matching and object recognition algorithm [63]. On the other hand, to obtain the Z position (focus distance), a numerical self-focusing function was applied on the reconstructed images [25, 64, 65]. The approach was used to highlight spermatozoa anomalous behaviours. In particular, the cell in Fig. 12 is affected by a morphological anomaly, known as "bent tail", which causes the non-linear out of plane motion (Fig.12a-b). Moreover, using the holograms collected to track the spermatozoon motion, the phase map image of the cell can be reconstructed (Fig. 12c) and as consequence, the morphological defect inducing the anomalous out of plane path can be highlighted.

The tracking of multiple spermatozoa, moving on different focal plane, is illustrated in fig. 13. In particular, five different human spermatozoa are successfully tracked. It's worth noting that the sequence of holograms is acquired at a constant sample – microscope objective distance, allowing an in vitro analysis and subsequently a numerical approach allows the 3D volumetric field reconstruction.

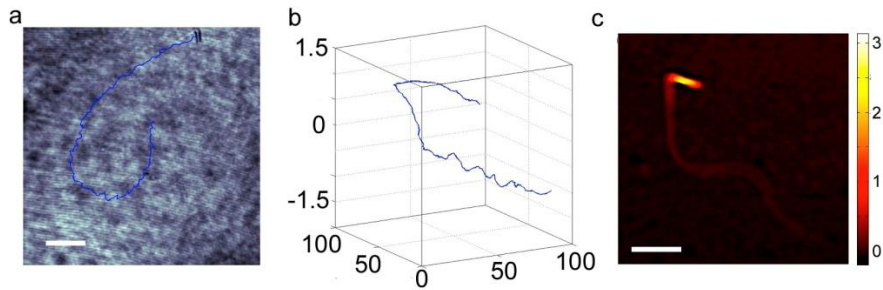

**Figure 12.** Single human sperm cells tracking. Transversal (a) and three-dimensional path (b) of a sperm cell presenting a bent tail; c) a phase map of the sperm cell, showing the morphological defect. Scale bar is 20 μm in (a) and 10 μm in (c), distances are reported in μm in (b); data were acquired over 36.8 s [62].

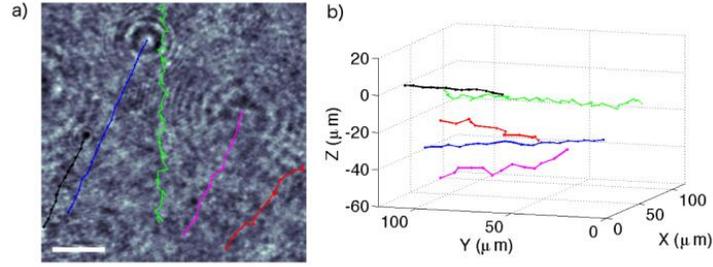

**Figure 13.** Multiple sperm cells tracking . Transversal (a) and reconstructed three-dimensional path (b). Scale bar is 20 µm, data were acquired over is 11 s. [62].

In the spermatozoa group shown in Fig. 13 an anomalous sperm cell is present, whose movement is plotted in green. In fact, while every other cell moves in parallel, swimming against a slight flow direction, the anomalous cell advances slower, along a broken track and on a tilted direction. Retrieving the motility parameters can highlight this anomalous behaviour (Table VI).

**Table VI.** Motility parameters evaluated for the cells plotted in Fig. 13. Rif [62]

|  | Curvilinear Velocity (VCL) [µm/sec] | Straight-line velocity (VSL) [µm /sec] | Linearity (VSL/VCL) [µm/µm] |
| --- | --- | --- | --- |
| Cell 1 (Green) | 16.9 | 9.8 | 0.58 |
| Cell 2 (Black) | 49.3 | 17.1 | 0.35 |
| Cell 3 (Purple) | 69.5 | 22.4 | 0.32 |
| Cell 4 (Red) | 90.0 | 20.2 | 0.22 |

In particular the VSL measured for Cell 1 (green) is lower than for every other cell and describes effectively the inefficient cell movement. Accordingly, the wobble is pretty uniform for Cell 2, 3, 4 and 5, varying between 0.97 and 0.99 and representing a motion with reduced oscillation around the average path. This value is instead sensibly lower for Cell 1, providing a quantitative description of the wide fluctuation of the spermatozoon head.

The possibility to retrieve the amount of spermatozoa into to a given volume also allows estimating the sperm cells concentration. Nowadays, this analysis, which is one of the first diagnostic analyses performed in order to evaluate clinic male infertility, is achieved by using a Makler chamber [66]; i.e. a collected part of the semen sample is placed into a calibrated volume. This approach is highly invasive. By applying the self-algorithm on only one acquired hologram, it should be possible to obtain the sperm concentration in a given volume.

## 4. Conclusion and Future trends

In this paper an overview on digital holography microscopy applied to both morphological and motility characterizations of sperm cells has been presented. Recent achievements obtained by means of digital holography have proved the possibility to provide three-dimensional information on both the morphology and motility of sperm. In fact, the great advantage to manage quantitative 3D information has allowed carrying out different numerical analysis (such as, volume, profiles along particular directions, selection of different zone) that has been used to better underline the differences between normal and abnormal sperm morphology. Moreover, the non-invasive feature of the DH approach has allowed the 3D tracking of the spatial motion of live spermatozoa in order to individuate the right movements of normal spermatozoa. All reported data show that the possibility to add the third dimension in the sperm analysis will provide a better statistic useful both to relate the sperm anomalies with male infertility and to enable differentiation of the healthier spermatozoa, thus enabling a non-invasive sterility examination using DHM.

Quantitative phase microscopy is a valuable optical assay for studies on measurement of cellular dry mass and its spatial dynamics [67]. Possible future developments of this study might consist in performing protein dry mass measurements on spermatozoa to evaluate the DNA content. Moreover, simple and cost-effective methods to convert existing microscopes in holographic imaging apparatus have been recently developed [68, 69]. We hope that these studies could facilitate the expansion of DHM and stimulate other researchers in the field of sperm cells biology to consider using DHM for their characterization studies.